\documentclass[12pt]{iopart} 
\usepackage{iopams,setstack}
\usepackage{graphicx}
\usepackage{dcolumn}

\newcommand{\ti}{\i}

\newcommand{\tro}{\"o}

\newcommand{\Sch}{Schr\tro dinger equation~}

\newcommand{\del}{$\delta$~}

\begin{document}

\title[BEC in a Linear Trap with a Dimple]{Bose-Einstein Condensate in a Linear Trap With a Dimple Potential}

\author{Haydar Uncu$^1$, Devrim Tarhan$^2$}
\address{$^{1}$Department of Physics, Adnan Menderes University,
Aytepe, 09100, Ayd\ti n, Turkey}
\address{$^2$Department of Physics, Harran University, Osmanbey
Yerle\c{s}kesi, 63300, \c{S}anl\i urfa, Turkey}
\date{\today}
\eads{huncu@adu.edu.tr,dtarhan@harran.edu.tr}
\begin{abstract}
We study Bose-Einstein condensation in a linear trap with a dimple
potential where we model dimple potentials by Dirac \del function.
Attractive and repulsive dimple potentials are taken into account.
This model allows simple, explicit numerical and analytical
investigations of noninteracting gases. Thus, the \Sch is used
instead of the Gross-Pitaevski equation. We calculate the atomic
density, the chemical potential, the critical temperature and the
condensate fraction. The role of the relative depth of the dimple
potential with respect to the linear trap in large condensate
formation at enhanced temperatures is clearly revealed. Moreover, we
also present a semi-classical method for calculating various
quantities such as entropy analytically. Moreover, we compare the
results of this paper with the results of a previous paper in which
the harmonic trap with a dimple potential in 1D was investigated.
\end{abstract}
%
\maketitle

\section{Introduction}
The experimental realization of Bose-Einstein condensates (BEC) of
dilute gases of alkalis $^{87}Rb$, $^{23}Na$ and $^7Li$ were
observed at very low temperatures following the improvement in
cooling and trapping techniques \cite{anderson,davis,bradley}. A
finite number of atoms $N$ which are confined in spatially
inhomogeneous trapping potentials are possible experimentally.
Despite the enhanced importance of phase fluctuations
\cite{khawaja}, several studies \cite{ketterle,druten} showed that
BEC can occur in harmonically trapped lower-dimensional systems for
finite $N$. Bose-Einstein condensation in one dimension (1D) with a
harmonic trap is attractive due to the enhanced critical temperature
and the condensate fraction \cite{ketterle,druten}. One- and two-
dimensional BECs were created in experiments\cite{gorlitz}.
One-dimensional quasicondensates were also generated on a microchip
\cite{ott,hansel} and in lithium mixtures \cite{schreck} a decade
ago.

The possibility of using all optical power-law traps have been
proposed for fast and efficient production of Bose-Einstein
condensates very recently\cite{bruce}. Taking into account weakly
interacting Bose gases, theoretical and experimental investigation
of momentum distribution of weakly interacting one-dimensional (1D)
Bose gases at the quasicondensation crossover have been studied
\cite{jacqmin}.

Using the semiclassical approximation, one can show that
Bose-Einstein condensation in one dimensional harmonic traps are not
stable\cite{yukalov}. Indeed, in all (1D) BEC experiments with
$^{87}Rb$, $^{23}Na$ and $^7Li$, the (1D) objects resembling BEC
that have been created are "quasi-condensates", which have
suppressed density fluctuations where the phase still fluctuates
along the cloud, and no single quantum state is macroscopically
occupied. In all of these experiments, the phase coherence length is
much smaller than the size of the cloud. The transition to the
quasicondensate regime is governed by interactions. (see Ref.
\cite{armijo} and references therein). The conditions for having
interaction induced quasicondensation versus finite-size
condensation (1D BEC formation) have been studied precisely
\cite{bouchoule}. In typical experiments (and actually all cases so
far), interactions are too strong to allow for the (1D) BEC instead
of the quasicondensate.

Cavalcanti et.al. showed even 3D gases in the presence of a uniform
field do not go undergo Bose-Einstein condensation at finite
temperatures \cite{cavalcanti}. However, the authors have revealed
that if there is a point-like impurity at the bottom of the vessel,
Bose-Einstein condensation occurs at $T\neq 0$. Moreover, as
mentioned in a previous paper, modification of the shape of the
trapping potential can be used to increase the phase space density
\cite{pinkse}. ``Dimple"-type potentials are the most favorable
potentials for this purpose. The phase-space density can be enhanced
by an arbitrary factor by using a dimple potential at the
equilibrium point of the harmonic trapping potential \cite{utem,
kurn}. The formation of a Bose-Einstein condensate in a cigar-shaped
three-dimensional harmonic trap, induced by the controlled addition
of an attractive "dimple" potential along the $z$ axis has been
realized very recently \cite{garrett}. Moreover, the effect of
variations of the longitudinal trap frequency on the BEC profile has
been analytically treated by using delta kicks \cite{ranjani} and
the ground state properties of 1D quantum gases in a harmonic trap
in the presence of a point-like attractive potential have been also
investigated \cite{busch}.

A simple magnetic trap mechanism is called quadrupole trap which
varies linearly in all directions and vanishes at the center. This
trapping mechanism has been used to obtain BEC of dilute gases
however it has a major disadvantage that there can be appreciable
losses in the vicinity of the node in the field \cite{dilute}.
There have been attempts to overcome  this disadvantage of the
simple quadrupole trap \cite{davis,dilute}. Dimple type potentials
are good candidates for this reason. Therefore we study the
properties of a BEC in a V shaped potential with a dimple at the
center. Attractive and repulsive dimple potentials are taken into
account. We model the dimple potential using a Dirac \del
function. The delta function can be defined via a Gaussian
function \cite{lighthill} $g(x,a)=(1/\sqrt{\pi}
a)\exp{(-x^2/a^2)}$ of infinitely narrow width $a$ so that
$g(x,a)\rightarrow\delta(x)$ for $a\rightarrow 0$. This allows for
analytical calculations in some limiting cases as well as a
simpler numerical treatment for arbitrary parameters. The \Sch for
a particle in the V-shaped potential decorated by a repulsive or
attractive Dirac delta function interaction at the center has been
solved very recently \cite{wangxin}. We apply this potential as a
model for quadrupole trap with a dimple.  We also compare the
results of this paper with the results of a previous paper in
which the harmonic trap with a dimple potential in 1D was
investigated \cite{utem}. On one hand, the linear trap could be
imagined as a disadvantageous against harmonic trap applications.
However, experimental easiness of may lead experimentalists to
prefer quadrupole trap with a dimple.

We calculate the transition temperature as well as the chemical
potential and condensate fraction for various number of atoms and
for various relative depths of the dimple potential. For describing
a system with interacting particles, the Gross-Pitaevski equation is
usually utilized. We note that we neglect the interactions between
the atoms in our model, and thus the \Sch for the linear trap with
the dimple potential is solved.

The paper is organized as follows. In Sect. II,  we present the
analytical solutions of the \Sch for a Dirac $\delta$-decorated
linear potential and the corresponding eigenvalue equation. In Sect.
III, we present a semiclassical approach and calculate the entropy
of a Bose gas in a V shaped potential. In Sec. IV, determining the
eigenvalues numerically, we show the effect of the dimple potential
on the condensate fraction, the transition temperature and the
chemical potential. Finally, we present our conclusions in Sect. V.

\section{Linear confining potential decorated with a Dirac \del potential at the origin}

We begin our discussion with the one dimensional linear potential
decorated with a Dirac \del function at x=0, ($\delta (x)$) . This
potential is given as:
\begin{equation}
V(x)=f \vert x \vert-\frac{\hbar^2}{2m} \sigma \delta(x),
\label{potential}
\end{equation}
where $m$ is the mass of the particles, $f$ is force term due to the
linear trap and $\sigma$ is the strength (depth) of the dimple
potential located at $x=0$. The factor $\hbar^2/2m$ is used for
calculational convenience. The negative $\sigma$ values represent
repulsive interaction while the positive $\sigma$ values represent
attractive interaction. The time-independent \Sch for the potential
given in Equation \eref{potential} is:
\begin{equation}
-\frac{\hbar^2}{2m} \frac{d^2 \Psi(x)}{dx^2} + V(x) \Psi(x)= E
\Psi(x). \label{Scheq}
\end{equation}
Although the solution of this equation can be found easily we
summarize it for the sake of completeness. First, we solve Equation
\eref{Scheq} for $x \neq 0$. Since the potential is an even
function, the eigenstates of this potential are even and odd
functions (see e.g. \cite{landau}). Hence it is enough to find the
solution for positive $x$. Defining $k^2=2 m E/\hbar^2$,
$l=\left(\hbar^2 /(2 m f)\right)^{1/3}$ and $u= x/l-k^2l^2$ for
$x>0$ and inserting these definitions into Equation \eref{Scheq} one
gets the Airy differential equation for the variable $u$
\cite{schwinger}:
\begin{equation}
\frac{d^2 \Psi(u)}{du^2}-u \,\Psi(u) =0 \label{Airy}
\end{equation}
Therefore, utilizing the fact that the wave functions are either
even or odd, one finds the wave functions in terms of Airy Ai
function $Ai(x) $  as {\cite{schwinger} \footnote{Since the second
solution of the Equation \eref{Scheq}, AiryBi function $Bi(x)$,
diverges for $x \rightarrow \infty$ the wave functions are expressed
only in terms of Airy Ai function}:
\begin{eqnarray}
\Psi_e(x)&=&c^{e}_n \, Ai[\frac{1}{l}(\vert x \vert-k_n^2 l^3)]
\quad \rm{~for~} n=0,2,\ldots \nonumber \\
\Psi_o(x)&=& c^{o}_n sgn(x) Ai[\frac{1}{l}(\vert x \vert-k_n^2 l^3
)] \quad \rm{~for~} n=1,3,\ldots  \; .\label{wavefunctions}
\end{eqnarray}
for $\sigma = 0 $ in Equation \eref{potential}. Here,  $c^{e}_n$ and
$c^{o}_n$ are the normalization constants for even and odd wave
functions respectively and $sgn(x)$ stands for the sign function.
The $c^{e}_n$ and $c^{o}_n$ can be found using the roots of $Ai(x)$
and the derivative of AiryAi function $Ai'(x)$ \cite{schwinger}.
Both $Ai(x)$ and $Ai'(x)$ have infinite number of zeros and all the
roots of these functions are negative. Let us denote the roots of
$Ai(x)$ by $-s_n$ and the roots of $Ai'(x)$ by $-z_n$, i.e
$Ai(-s_n)=Ai'(-z_n)=0$. These coefficients can be ordered as $
-z_1<-s_1<-z_2<-s_2 < \ldots$ . Ordering these coefficients just
using one symbol like $\alpha_0=z_1, \alpha_1=s_1, \ldots
\alpha_{2n-1}=s_n, \alpha_{2n-2}=z_n, \; n=1,2,\ldots $, one gets
for the normalization coefficients \cite{schwinger}:
\begin{eqnarray}
c^{e}_n &=& \left[2 \alpha_n l \left[ Ai(-\alpha_n) \right]^2\right]^{-1/2}
\quad \rm{~for~} n=0,2,\ldots \nonumber\\
c^{o}_n &=&  \left[ 2 l Ai'(-\alpha_n)  \right]^{-1}  \quad
\rm{~for~} n=1,3,\ldots \; . \label{roots}
\end{eqnarray}
Using the fact that $\Psi_o(0)=0$ and $(d \Psi_e(x)/dx)_{x=0}=0$,
one finds the eigenvalues of the Hamiltonian whose the potential
term is given in Equation \eref{potential} for $\sigma =0 $:
\begin{equation}
E_n=\left( \frac{\hbar^2 f^2}{2m}\right)^{1/3}   \alpha_n \quad
\rm{for~} n=0,1,\ldots \; .\label{eigen0}
\end{equation}
We define
\begin{equation}
k_n=\sqrt{(2mE_n)}/\hbar .\label{knvalues}
\end{equation}
Using $l=\left(\hbar^2 /(2 m f)\right)^{1/3}$ we get
\begin{equation}
k_n=\sqrt{ \alpha_n}/l \quad \rm{for~} n=0,1,\ldots \;
.\label{eigen0_1}
\end{equation}
If $\sigma$ in Equation \eref{potential} is not zero, the
eigenvalues of the even states can be found by taking the integral
of the Schr\"odinger equation from $-\epsilon$ to $\epsilon$ in the
limit $\epsilon \rightarrow 0$:
\begin{eqnarray}
&\lim_{\epsilon \to 0 }&  \left( -\frac{\hbar^2}{2m}
\int\limits^{\epsilon}_{-\epsilon} \frac{d^2 \Psi(x)}{dx^2}  dx + f
\int\limits^{\epsilon}_{-\epsilon} \vert x \vert \Psi(x) dx -
\frac{\hbar^2}{2m} \sigma \int\limits^{\epsilon}_{-\epsilon}
\delta(x) \Psi(x) dx \right) = \nonumber \\& \lim_{\epsilon \to 0 }&
\, E \int\limits^{\epsilon}_{-\epsilon}  \Psi(x) dx . \label{limit1}
\end{eqnarray}
Since the functions  $\vert x \vert \psi(x)$ and $\psi(x)$ are
continuous at $x =0$ the second integral on the left hand side and
the integral on the right hand side of the Equation \eref{limit1}
vanish. Therefore this equation reduces to
\begin{equation}
\lim_{\epsilon \to 0 } \frac{d \psi(x)}{dx}\vert_{x=\epsilon}-
\frac{d \psi(x)}{dx}\vert_{x=\epsilon} + \sigma \psi(0)=0
\label{limit2}
\end{equation}
This equation reveals that the derivative of the even eigenfunctions
are not continuous when $\sigma \neq 0 $. Therefore, the eigenvalues
of the even eigenfunctions cannot be found from the roots of
$Ai'(x)$ in this case. However, substituting $\Psi_e(x)$ in Equation
\eref{wavefunctions} for the $\psi(x)$ into the Equation
\eref{limit2} we get an equation for the eigenvalues of the even
eigenfunctions for the potential given in Equation \eref{potential}
for $\sigma \neq 0 $:
\begin{equation}
 \frac{2 Ai'(-k_n^2 l^2)}{Ai(-k_n^2 l^2)}=- \sigma l .
 \label{eigenvaleq}
\end{equation}
This equation can be solved numerically for $k_n$.  Substituting
these $k_n$ values into the argument of AiryAi function for even
wave functions in Equation \eref{wavefunctions} one gets the even
eigenfunctions for $\sigma \neq 0$ of the Hamiltonian for the
potential given in Equation \eref{potential}.

The odd wave functions vanish at $x=0$. Hence their derivatives are
continuous and they are not affected by Dirac $\delta$ function at
the origin. Therefore the eigenvalues corresponding to the odd
eigenfunctions of the potential in Equation \eref{potential} are the
same with the eigenvalues of odd eigenfunctions of the potential
$V(x)=f \vert x \vert$. On the other hand, we showed that the energy
eigenvalues of even states change as a function of $\sigma$. As
shown in Ref. \cite{wangxin}, the ground state energy eigenvalue
decreases without a lower limit as $\sigma$ increases (attractive
case). However, the energies of the excited even states are limited
by the energies $E_{2n+1}$ of odd states and as $\sigma \rightarrow
\infty$, $E_{2n+2} \rightarrow E_{2n+1}$ where $n=0,1,...$.

\section{Semiclassical Approach}

We will first investigate Bose gas in a V shaped potential with a
Dirac $\delta$ at the origin semiclassically \footnote{We propose
that this potential provides a model for a quadrupole trap with a
dimple}. The spectrum of the Hamiltonian with the potential in
Equation \eref{potential} is discrete for all values of $\sigma$
including $\sigma=0$ (only the linear confining potential). For
convenience, we  define a dimensionless parameter $\Lambda$ in terms
of $\sigma$ as:
\begin{equation}
\Lambda= \sigma l . \label{lambda}
\end{equation}
We choose the value of $l$ such that $ \Lambda$ defined in equation
above has the same order of magnitude with $\Lambda$ values defined
in Ref. \cite{utem} for a given $\sigma $, so that we can compare
the effect of dimple potential to the linear confining potential
with the effect of dimple potential to the the harmonic trap. In
\cite{utem} the calculations are made for $^{23}Na$ particles with
$m=23$ amu and the frequency $\omega$ of the harmonic trap was
chosen as $\omega= 2 \pi 21 Hz$. Equating the length scales of the
harmonic trap $\sqrt{\hbar/(2 m \omega)}$ to the length scale of the
confining linear potential $l$, we get for $l=3.23 \mu m$.

Quantum mechanically - as we will present in the following section-
one calculates the thermodynamic variables of a non-interacting
condensate like critical temperature, condensate fraction and the
chemical potential taking sums over the discrete energy eigenvalues.
However, in the semiclassical approximation one uses the density of
states and employs integrals instead of sums (see e.g.
\cite{dilute,stringari}).  Therefore we first find the density of
states for a linear confining potential. As we have showed in the
previous section, the odd and even energy eigenvalues of the linear
confining potential ($\sigma=0$ in Equation \eref{potential}) is
proportional to the negative of the root values of $Ai(x)$ and to
the $Ai'(x)$ ($s_n$ and $z_n$ values defined in the previous
section), respectively. This proportionality can be used to
calculate the density of states for linear confining potential. The
$s_n$ values can be approximated by the formula \cite{schwinger},
\footnote{This approximation can be improved to get a better
accuracy \cite{schwinger}. However, since we are only interested in
the density of states and more accurate formulae give the same
density of states we use the simplest approximation to the roots.}:
\begin{equation}
-s_n \approx [\frac{3 \pi}{2} (n-\frac{1}{4})]^{2/3} .
\label{Airyrootapprox}
\end{equation}
$-s_n$ values are an approximation to the roots of $Ai$. Using the
definitions $\alpha_n$, $s_n$ and Equation \eref{eigen0}, we get an
equality between $s_n$ and the eigenenergies of odd states as
$-s_n=(2 m/\hbar^2 f^2)^{1/3} E_n$. We take $n$ and therefore
$E_n\equiv E$ as continuous variables and get $\frac{3 \pi}{2}
(n-\frac{1}{4}) = (2 m/\hbar^2 f^2)^{1/2} E^{3/2}$. So we obtain the
density of the odd states
\begin{equation}
\frac{dn}{dE}=\frac{1}{ \pi} (\frac{2m}{\hbar^2 f^2})^{1/2} E^{1/2}.
\label{dosostate}
\end{equation}
For each odd state there is an even state. Hence the density of
states ($g(E)$) for the linear confining potential is
\begin{equation}
g(E)=\frac{2}{ \pi} (\frac{2m}{\hbar^2 f^2})^{1/2} E^{1/2}.
\label{dos}
\end{equation}

Using the density of states one can calculate the number of
particles $N_T$ in the thermal gas that is the number of particles
which are not in  condensate phase \cite{dilute}. Above the critical
temperature $T_c$ the number of particles in the ground state is
negligible and all the particles are assumed to be in the thermal
gas. As the temperature decreases the chemical potential increases
and comes very close to the ground state energy. Therefore one gets
the critical temperature taking the chemical potential $\mu$ equal
to the ground state value \cite{stringari}\footnote{As mentioned in
Ref. \cite{ketterle}, the phase transitions due to discontinuity in
an observable macro parameter occurs only in thermodynamic limit,
where $N \rightarrow \infty$. Therefore we assume the critical
temperature $T_c$ is the temperature that particles begin to
accumulate in the ground state.}. Using the density of states and
Bose distribution we find:
\begin{equation}
N=\int_{\epsilon_1}^{\infty} \frac{g(\epsilon)}{e^{\beta_c
(\epsilon-\mu)-1}} d\epsilon=\frac{2}{\pi} (\frac{2 m}{\hbar^2
f^2})^{1/2} \int_{\epsilon_1}^{\infty}
\frac{\epsilon^{1/2}}{e^{\beta_c (\epsilon-\mu)-1}} d\epsilon
\label{semiclTc}
\end{equation}
where the lower limit of the integral is the energy eigenvalue of
the first excited state and $\beta_c=1/(k_B T_c)$. For different
$\Lambda $ values which are used to model the strength of the dimple
potential we can find the energy eigenvalue of the ground state
using Equation \eref{eigenvaleq}. The energy eigenvalue of the first
excited state is same for all $\Lambda$ and found by Equation
\eref{eigen0} taking $n=1$ because it is an odd eigenfunction. The
Dirac \del potential in Equation \eref{potential} does not change
the density of states. That is it does not create or destroy
eigenstates but only change the values of the energy eigenvalues.
Therefore the same density of state expression can be used for all
$\Lambda$ values. However, the total number of particles in the
thermal gas or (in the condensate) differ for different  $\Lambda$
values because as $\Lambda$ varies  the energy eigenvalue of the
ground state varies which cause a change in the chemical potential
value.

So, if the Equation \eref{semiclTc} can be solved for $\beta_c$ for
fixed N, one can find the change of critical temperature with
increasing $\sigma $ which determines the strength of the dimple
potential. We approximately calculate the integral in Equation
\eref{semiclTc} taking the lower limit zero (i.e we set
$\epsilon_1=0$) and shifting the chemical potential by the same
amount ($\mu \rightarrow \mu-\epsilon_1$). Applying these
approximations we take the integral in Equation \eref{semiclTc} and
find
\begin{equation}
N \approx 2 (\frac{2 m}{\pi \hbar^2 f^2})^{1/2}
(\frac{1}{\beta_c})^{3/2} g_{3/2}(e^{\beta_c \mu}).
\label{semiclTcapprox}
\end{equation}
where the Bose function $g_{p}(x)$ is defined as \cite{dilute}:
\begin{equation}
g_{p}(x)=\sum_{l=1}^{\infty} \frac{x^l}{l^{p}} \, .
\label{bosefunctions}
\end{equation}
We have solved Equation \eref{semiclTcapprox} and find $T_c$ for
various $\Lambda$ values. We will present our results in the
following section in order to compare the results obtained from
semiclassical approximation with the results obtained from quantum
mechanical calculations.

Adding a dimple potential adiabatically changes the critical
temperature by changing the value of chemical potential $\mu$
\cite{stringari}. If an attractive dimple is added to the trap
potential at temperatures slightly above the critical temperature of
the trap without a dimple adiabatically, the entropy remains
constant. Hence the decrease in the ground state energy cause the
formation of BEC \cite{stringari}. In the laboratory this process
can be maintained adiabatically and the entropy of the boson gas
remains constant during the process. Therefore if one finds an
expression for the entropy of the Bose gas one can calculate the
value of condensate fraction of a BEC obtained for an adiabatic
process.

The entropy of a non-interacting Bose gas is easily calculated using
the grand potential $\Omega$ \cite{stringari}:
\begin{equation}
\Omega= \Omega_0 + k_B T \sum_i ln[1-e^{\beta (\mu-\epsilon_i)}] .
\label{defgrand}
\end{equation}
Using the density of states for a linear confining potential given
in Equation \eref{dos}, the grand potential of a non-interacting
Bose gas trapped using linear confining potential can be written as
\begin{equation}
\Omega= \Omega_0 + \frac{2 k_B T}{\pi} (\frac{2 m}{\hbar^2
f^2})^{1/2} \int_{\epsilon_1}^{\infty}  \epsilon^{1/2}
\ln[1-e^{\beta (\mu-\epsilon)}] \, \rmd \epsilon.
\label{lineargrand}
\end{equation}
Calculating the integral in Equation \eref{lineargrand} one finds
\begin{eqnarray}
&\Omega = \Omega_0 - \frac{2}{\pi} (\frac{2 m}{\hbar^2 f^2})^{1/2}
\times \nonumber \\ &  \left[ \frac{\sqrt{\epsilon_1}}{\beta^2}
g_2(e^{\beta (\mu-\epsilon_1)})+ \frac{\sqrt{\pi}}{2 \beta^{5/2}}
\sum_n^{\infty} \frac{e^{n \beta \mu} \rm{Erfc} (\sqrt{n \beta
\epsilon_1})}{n^{5/2}} \right]
\label{lineargrandcalculated}
\end{eqnarray}
where  $\beta=1/(k_B T)$,
$\rm{Erfc}(x)=1-\rm{Erf}(x)=1-(2/\sqrt{\pi})\int_0^x e^{-t^2} dt $
and $g_p(x)$ is defined in \eref{bosefunctions}. Then, we find the
entropy by $S=-d{\Omega}/{dT}$:
\begin{eqnarray}
&S =k_B \frac{2}{\pi} (\frac{2 m}{\hbar^2 f^2})^{1/2} \times
\nonumber \\ & \bigg{\{}  \frac{5 \sqrt{\epsilon_1}}{2 \beta}
g_2(e^{\beta (\mu-\epsilon_1)})+ \sqrt{\epsilon_1} (\mu-\epsilon_1)
\ln[1-e^{\beta (\mu-\epsilon_1)}]   + \nonumber   \\ & \frac{5
\sqrt{\pi}}{2 \beta^{3/2}} \sum_{n=1}^{\infty} (\frac{e^{n \beta
\mu} \rm{Erfc} (\sqrt{n \beta \epsilon_1})}{n^{5/2}}) -
\frac{\sqrt{\pi} \mu}{2 \beta^{1/2}} \sum_{n=1}^{\infty} \frac{e^{n
\beta
\mu} \rm{Erfc} (\sqrt{n \beta \epsilon_1})}{n^{3/2}}  \bigg{\}}. \nonumber \\
\label{entropy}
\end{eqnarray}
For the temperatures slightly above $T_c$, one can take
$\mu=\epsilon_0$ and $\beta=\beta_c$. Therefore for temperatures
slightly above $T_c$, Equation\eref{entropy} becomes
\begin{eqnarray}
&S =k_B \frac{2}{\pi} (\frac{2 m}{\hbar^2 f^2})^{1/2} \times
\nonumber \\ & \bigg{ \{ }  \frac{5 \sqrt{\epsilon_1}}{2 \beta_c}
g_2(e^{\beta_c (\epsilon_0-\epsilon_1)})+ \sqrt{\epsilon_1}
(\epsilon_0-\epsilon_1) \ln[1-e^{\beta_c (\epsilon_0-\epsilon_1)}]
\nonumber
\\ & + \frac{5 \sqrt{\pi}}{2 \beta_c^{3/2}} \sum_{n=1}^{\infty}
(\frac{e^{n \beta_c \epsilon_0} \rm{Erfc} (\sqrt{n \beta_c
\epsilon_1})}{n^{5/2}})  \nonumber \\ &- \frac{\sqrt{\pi}
\epsilon_0}{2 \beta_c{1/2}} \sum_{n=1}^{\infty} \frac{e^{n \beta_c
\epsilon_0} \rm{Erfc} (\sqrt{n \beta_c \epsilon_1})}{n^{3/2}} \bigg{
\}}
\nonumber \\
\label{entropy2}
\end{eqnarray}
where $\epsilon_0=\left( \frac{\hbar^2 f^2}{2m})^{1/3}  \right) z_1$
and $ \epsilon_1=\left( \frac{\hbar^2 f^2}{2m})^{1/3} \right) s_1$.
Adding a dimple potential changes the ground state therefore the
chemical potential changes also because $\mu=\epsilon_g < \epsilon_0
$. So, we get
\begin{eqnarray}
&S =k_B \frac{2}{\pi} (\frac{2 m}{\hbar^2 f^2})^{1/2} \times
\nonumber \\ & \bigg{ \{ }  \frac{5 \sqrt{\epsilon_1}}{2 \beta}
g_2(e^{\beta_f (\epsilon_g-\epsilon_1)})+ \sqrt{\epsilon_1}
(\epsilon_g-\epsilon_1) \ln[1-e^{\beta_f (\mu-\epsilon_1)}]
\nonumber \\ &  + \frac{5 \sqrt{\pi}}{2 \beta_f^{3/2}}
\sum_{n=1}^{\infty} (\frac{e^{n \beta_f \epsilon_g} \rm{Erfc}
(\sqrt{n \beta_f \epsilon_1})}{n^{5/2}}) \nonumber \\ & -
\frac{\sqrt{\pi} \epsilon_0}{2 \beta_f{1/2}} \sum_n^{\infty}
\frac{e^{n \beta_f \epsilon_g} \rm{Erfc} (\sqrt{n \beta_f
\epsilon_1})}{n^{3/2}} \bigg{ \} }.
\nonumber \\
\label{entropy3}
\end{eqnarray}
for the entropy of the BEC  in the linear trap with a dimple
potential where $\beta_f=1/(k_b T_f)$ and $T_f$ is the final
temperature of the system. Since we model the dimple potential with
a Dirac \del potential, we can calculate the new ground state energy
($\epsilon_g$), -i.e the ground state energy of the trap with a
dimple- using Equation \eref{eigenvaleq}. Then, we equate Equation
(\ref{entropy2}) and Equation (\ref{entropy3}) to calculate the
final temperature of the system, because the entropy remains
constant during adiabatic process. Then, we calculate condensate
fraction for this final temperature.

We present the increase of the condensate fraction with the decrease
in the ground state energy value in Fig. (\ref{compCF}). To obtain
this figure, we have first calculated the ground state energy values
($\epsilon_g$'s) of linear confining potential with a dimple
potential for different strengths of the dimple i.e for different
$\sigma$ ($\Lambda$) values in our model. Then applying the
procedure outlined in the previous paragraph we find $T_f$. Finally,
we use Equation \eref{semiclTcapprox} to find the number of
particles in the thermal gas $N_T$. Since we assume that the number
of particles $N$ in the Bose gas  is fixed we get the condensate
fraction using ($N_0/N=1-N_T/N$). We choose the range of $\sigma$
values which determines the strength of the dimple same as in the
paper \cite{utem}, i.e. between $ 0 \, \mathrm{m}^{-1} $ and $10^{8}
\, \mathrm{m}^{-1} $.
\begin{figure}
\centering{\vspace{0.5cm}}
\includegraphics[width=3.5in]{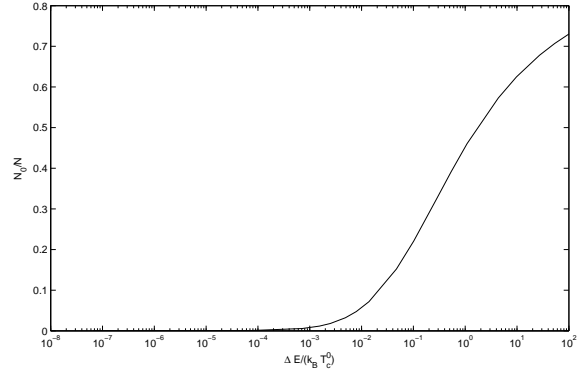}
\caption{The condensate fraction values evaluated with the
semiclassical method  for $N=10^4$.  The logarithmic scale is used
for x axis, where $\Delta E=  \epsilon_0 -\epsilon_g$.}
\label{compCF} \end{figure}

\section{BEC In a One-Dimensional Linear Potential with a Dirac \del Function}

In this section, we will investigate the behavior of a  BEC confined
in a one-dimensional linear confining potential decorated with a
Dirac \del function given in Equation \eref{Scheq} quantum
mechanically. That is we will use the discrete energy eigenvalues
calculated numerically by Equation \eref{eigen0_1} and Equation
\eref{eigenvaleq} for odd ($n=1,3,\cdots$) and even ($n=0,2,\cdots$)
states, respectively.

First we will investigate the change of the critical temperature as
a function of $\Lambda$ which shows the effect of  the dimple
potential in our model.
\begin{figure}
\centering{\vspace{0.5cm}}
\includegraphics[height=2.5 in, width=3.5 in]{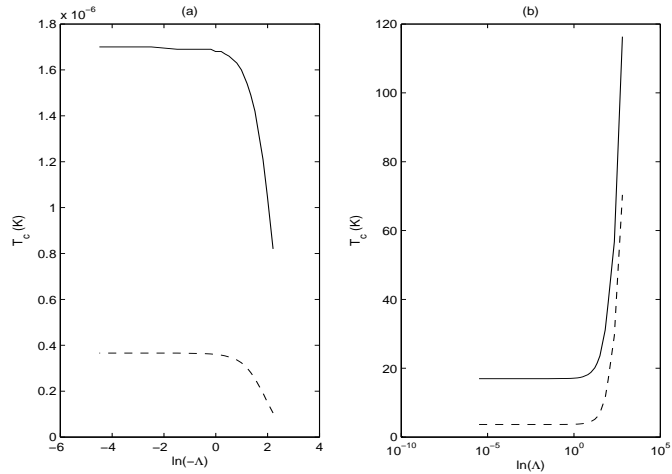}
\caption{(a) The critical temperature $T_c$  vs. $\Lambda $ for
$N=10^4$ and $N=10^5$ and for negative $\Lambda $. $\Lambda $ is a
dimensionless variable defined in Equation (\ref{lambda}). The solid
line and the dashed line show $T_c$ values with decraeasing
$\Lambda$ for $N=10^4$ and $N=10^5$, respectively. Here we use m
$=23$ amu ($^{23}\mathrm{Na}$) and $l=3.23 \mu m$. The logarithmic
scale is used for $-\Lambda$ axis. (b) The critical temperature
$T_c$ vs. $\Lambda $ for $N=10^4$ and $N=10^5$ and for
positive$\Lambda $. The logarithmic scale is used for $\Lambda$
axis.} \label{critemp}
\end{figure}
We will present the change of the critical temperature with respect
to $\Lambda$ both for attractive and repulsive dimple.  The critical
temperature ($T_c$) is obtained by taking the chemical potential
equal to the ground state energy ($\mu=E_g=E_0$) and
\begin{equation}
N \approx \sum_{i=1}^{\infty} \frac{1}{e^{\beta_c \varepsilon_i}-1}
\;\; , \label{Tcritical}
\end{equation}
at $T=T_c$, where $\beta_c=1/(k_B T_c)$. For finite $N$ values, we
define $T_c^0$ as the solution of Equation (\ref{Tcritical}) for
$\Lambda=0$ i.e. only for the linear confining potential.

In Equation (\ref{Tcritical}),  $ \varepsilon_i$'s are the
eigenvalues for the potential given in Equation (\ref{potential}).
For $\Lambda \neq 0$ the energies of odd states are unchanged and
equal to $\sqrt{\alpha_n}/l$ for $n=1,2,3$. The energies of even
states are found by solving Equation (\ref{eigenvaleq}) numerically.
Then, these values are substituted into Equation (\ref{Tcritical});
and finally this equation is solved numerically to find $T_c$.

For positive $ \Lambda $, we present the change of $T_c$ with
increasing $ \Lambda $ for $N=10^4$ and $N=10^5$ in Fig.
(\ref{critemp}-b). For negative $\Lambda$, we present the change
with increasing absolute $\Lambda$ in Figure (\ref{critemp}-a). In
this figure, logarithmic scale is used for $ \vert \Lambda \vert$
axes. As $\vert \Lambda \vert$ increases, the critical temperature
increases (decreases) very rapidly when $ \vert \Lambda \vert \sim
10^2$ for positive (negative) $ \vert \Lambda \vert$. However the
decrease of critical temperature  for negative $\Lambda$ is not as
rapid as the increase for positive $\Lambda$. This is due to the
fact that as $\vert \Lambda \vert$ increases for the attractive case
the ground state energy value decreases indefinitely however the
excited states are bounded. The limit value of the eigenenergies of
the excited states  with a tough dimple (modeled by a Dirac
$\delta$) are the eigenvalues of the one lower level of the
confining potential without a dimple. On the other hand for the
repulsive case (negative $\Lambda$ in our case) all the energies
with a tough dimple are bounded above by the eigenvalues of one
higher level of the confining potential without a dimple.
\begin{figure}
\centering{\vspace{0.5cm}}
\includegraphics[height=2.5 in, width=5.5 in]{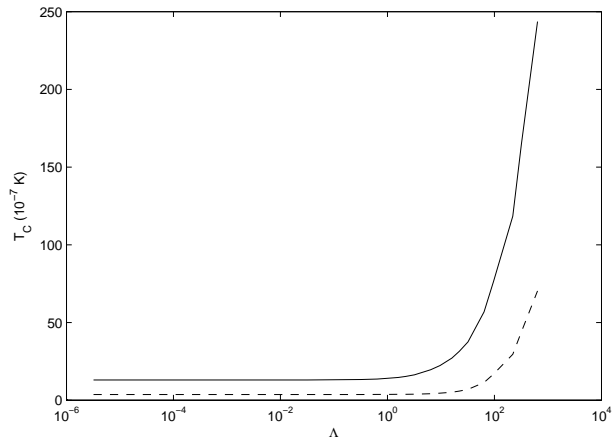}
\caption{The critical temperature $T_c$  vs. $\Lambda $ for $N=10^4$
for harmonic and linear trap with a dimple, respectively. $\Lambda $
is defined in Equation (\ref{lambda}). The solid line and the dashed
line show $T_c$ values with increasing $\Lambda$ for harmonic and
linear traps, respectively. The other parameters are same as Fig.
(\ref{critemp}) The logarithmic scale is used for $\Lambda$ axis.}
\label{figcomp}
\end{figure}

At this point it may be useful to compare the critical temperature
values of a harmonic trap and linear trap with in 1D with the
increasing depth of dimple ($\Lambda$ values in our model)
\footnote{The change of the critical temperature with respect to
increasing depth of dimple for a harmonic trap is presented in Ref.
\cite{utem}. }. We show the change of the critical temperature with
respect to $\Lambda$ for both cases in Fig. (\ref{figcomp}). The
critical temperature values are higher for harmonic trap compared to
linear trap if one chooses the length scales same for both
potential. This is an expected result because the confinement
strength of harmonic potential which increases quadratically with
respect to distance to the center of the potential is larger than
linear trap which increases linearly with respect to distance to the
center.

As we mentioned in the previous section the increase of the critical
temperature can be calculated approximately by solving Equation
\eref{semiclTc} obtained by the semi-classical method. We compare
the critical temperature values obtained quantum mechanically and
semiclassically in Figure (\ref{error}).  This figure shows that the
agreement of the results of two methods is better for large
$\Lambda$ (or large $\Delta E$) values. This is due to the fact that
the average occupations for excited states is so low ($ n^{excited}
\ll 1$) that the semiclassical treatment for this gas is valid.
Therefore, the semiclassical method may be useful for calculations
with large $\Lambda$ values because for $\Lambda \approx 10^3 $  the
numerical calculations do not give accurate results.

\begin{figure}
\centering{\vspace{0.5cm}}
\includegraphics[height=2.5 in,width=3.5 in]{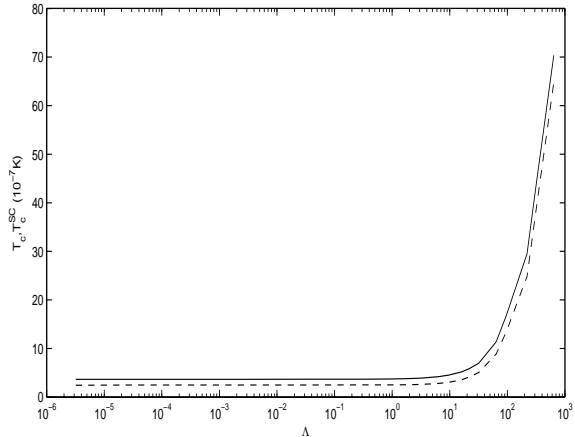}
\caption{ ($T_c^{SC}, \, T_c) $ vs. $\Lambda$ for $N=10^4$. The
dashed line shows the $T_c^{SC}$ values obtained semiclassically
where the solid line shows the $T_c$ values obtained quantum
mechanically. The logarithmic scale is used for $\Lambda$ axis.}
\label{error}
\end{figure}
For a gas of N identical bosons, the chemical potential $\mu$ is
obtained by solving equation
\begin{equation}
N=\sum_{i=0}^{\infty} \frac{1}{e^{\beta (\varepsilon_i - \mu)}-1}=
N_0 + \sum_{i=1}^{\infty} \frac{1}{e^{\beta (\varepsilon_i -
\mu)}-1}, \label{chem pot} \end{equation}
at constant temperature and for given N, where $\varepsilon_i$ is
the energy of state $i$. We present the change of $\mu $ as a
function of $T/T_c^0$ for $ N=10^4 $ in Fig.(\ref{muvstgraph}) and
in Fig.(\ref{muvstgraph2})  for $ \Lambda = 0$ and  $ \Lambda = 32
$, respectively.
\begin{figure}
\centering{\vspace{0.5cm}}
\includegraphics[width=3.5in]{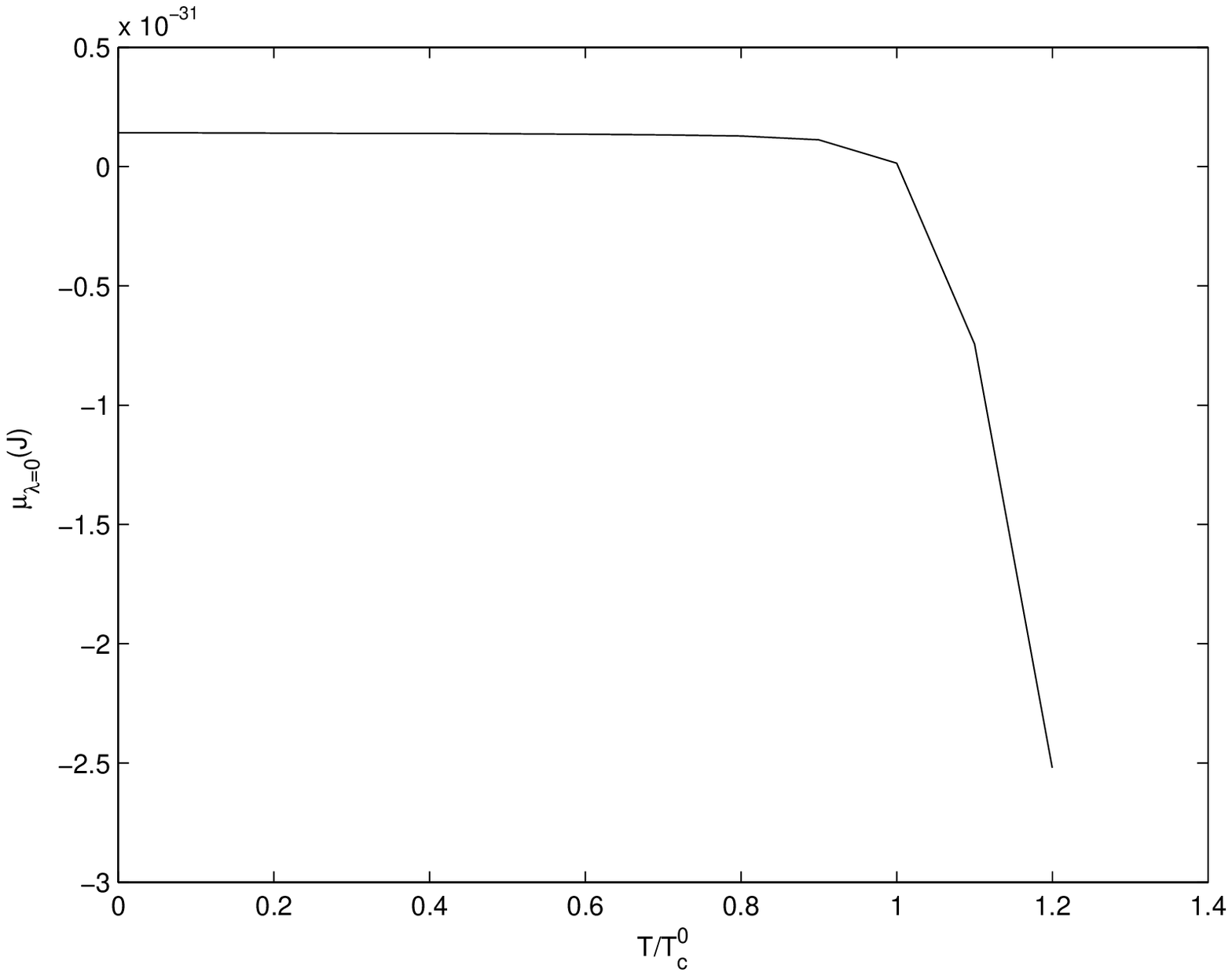}
\caption{The chemical potential $\mu$ vs temperature $T/T_c^0$ for
$N=10^4$ and  $\Lambda=0$. The other parameters are the same as Fig.
\ref{critemp}.}
 \label{muvstgraph}
\end{figure}
\begin{figure}
\centering{\vspace{0.5cm}}
\includegraphics[width=3.5in]{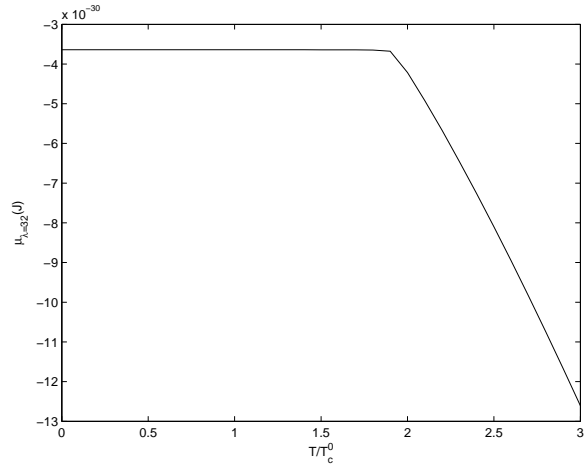}
\caption{The chemical potential $\mu$ vs temperature $T/T_c^0$ for
$N=10^4$ and $\Lambda=32$. The other parameters are the same as Fig.
\ref{critemp}.}
 \label{muvstgraph2}
\end{figure}
By inserting $\mu $ values into the equation
\begin{equation} N_0= \frac{1}{e^{\beta (\varepsilon_0 - \mu)}-1},
\end{equation}
we find the average number of particle in the ground state. $N_0/N$
versus $T/T_c^0$ for $N=10^4$, $\Lambda= 0,  3.2 , 32 $ are shown in
Fig. (\ref{condfrac}).
\begin{figure}
\centering{\vspace{0.5cm}}
\includegraphics[width=3.5in]{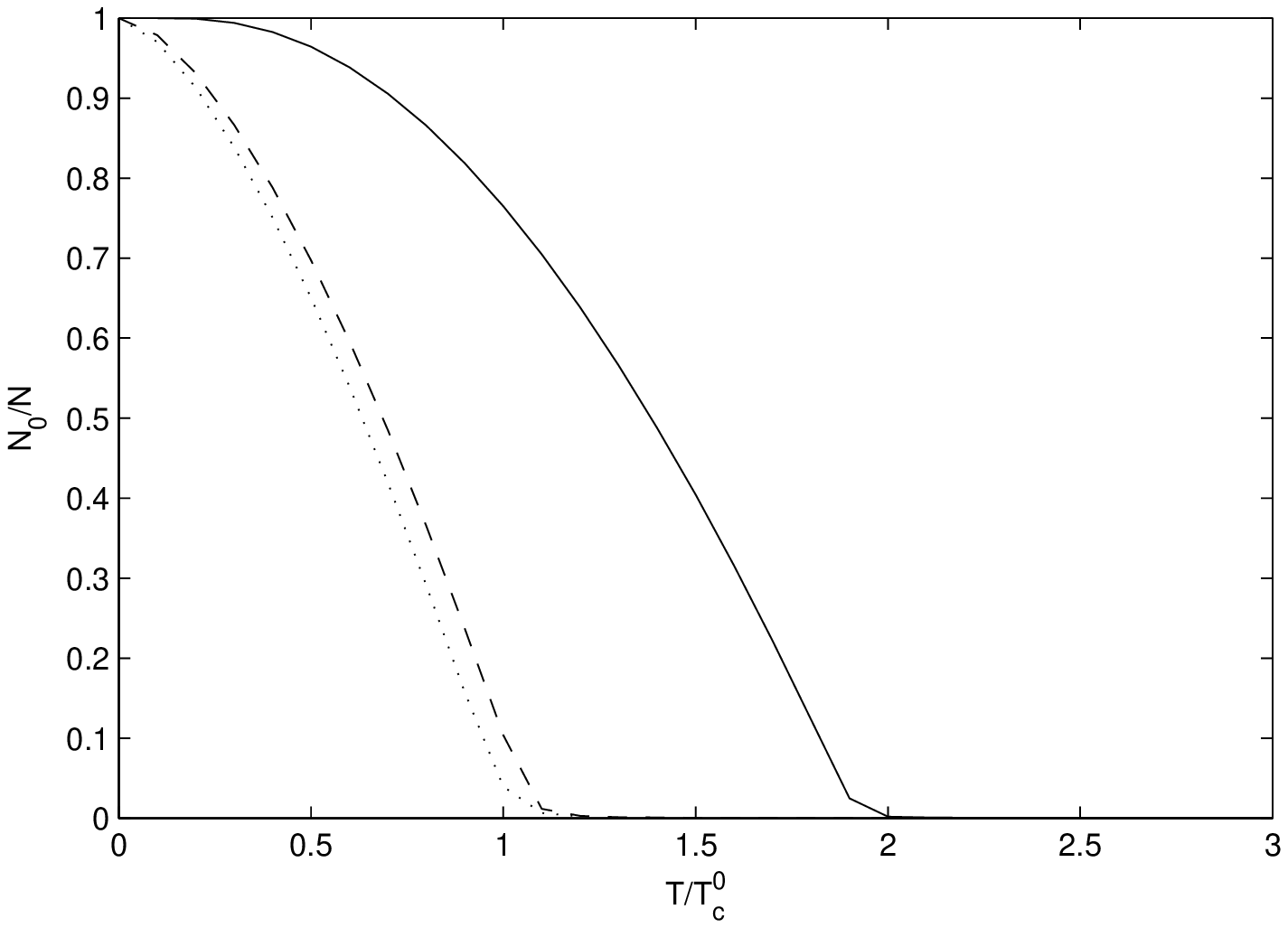}
\caption{$N_0/N$ vs $T/T_c^0$ for $N=10^4$  dotted line for
$\Lambda=0$, dashed line for $\Lambda=3.2$, solid line  for
$\Lambda=32$. The other parameters are the same as Fig.
\ref{critemp}.} \label{condfrac}
\end{figure}
We also present the change of condensate fraction for $N=10^4$ and
$N=10^5$ with increasing temperature for fixed $\Lambda =32$ in Fig.
(\ref{condfrac2}). The condensate fractions for different N values
are drawn by using their corresponding $T_c^0$ values. The $T_c^0$
values are $0.37\,\mu$K for $ N=10^4 $, and $1.7 \,\mu$K for $
N=10^5 $.
\begin{figure}
\centering{\vspace{0.5cm}}
\includegraphics[width=3.5in]{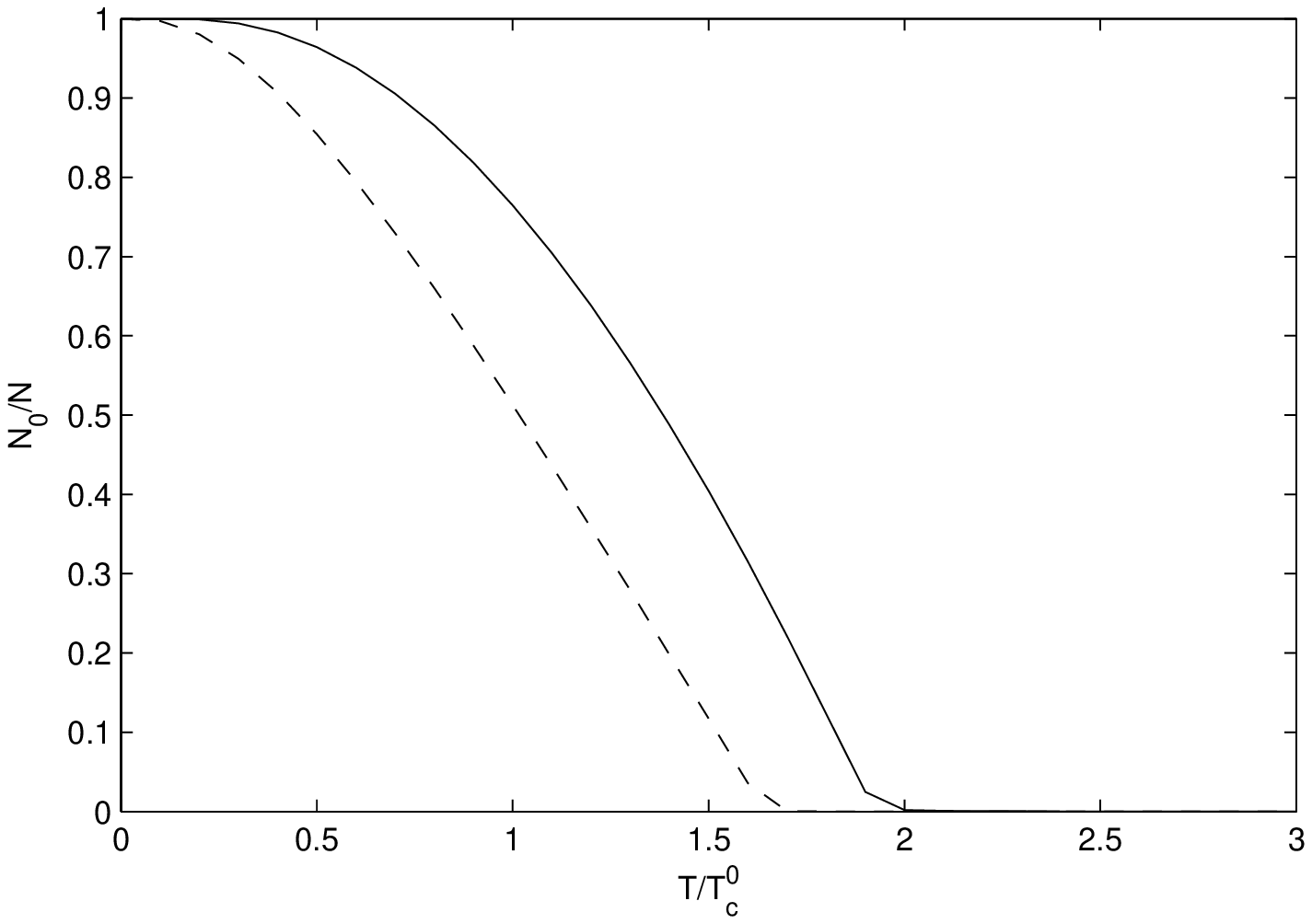}
\caption{Condensate fraction $N_0/N$ vs $T/T_c^0$ for fixed
$\Lambda=32$. The solid and dashed lines show condensate fraction
for $N=10^4$ and $N=10^5$, respectively. Logarithmic scale is used
for $\Lambda$ axis. The other parameters are the same as Fig.
\ref{critemp}.}
 \label{condfrac2}
\end{figure}
\begin{figure}
\centering{\vspace{0.5cm}}
\includegraphics[width=3.5in]{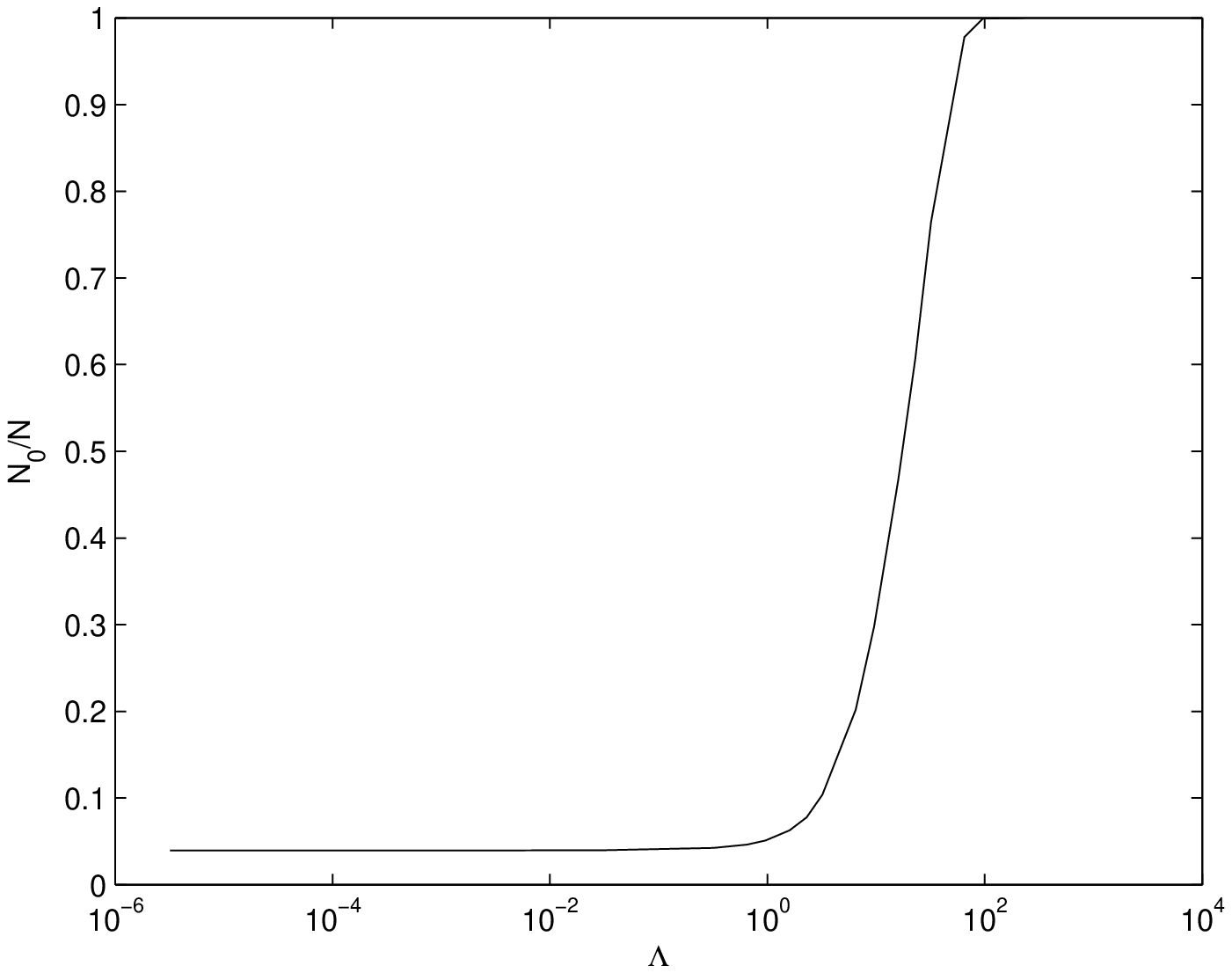}
\caption{Condensate fraction $N_0/N$ vs the strength of the Dirac
\del potential $\Lambda$ for $N=10^4$ at $T=T_c^0$. The logarithmic
scale is used for $\Lambda$ axis. The other parameters are the same
as Fig. \ref{critemp}.} \label{condfvss}
\end{figure}

We also find the condensate fraction as a function of  $\Lambda$ at
a constant $T=T_c^0$. These results for $N=10^4$ are shown in Fig.
(\ref{condfvss}). We notice that  large $\Lambda$ values
($\Lambda>1$) induce sharp increase in condensate fraction like in
the harmonic trap with a dimple potential \cite{utem}.

\begin{figure}[h!]
\centering{\vspace{0.5cm}}
\includegraphics[width=3.5in]{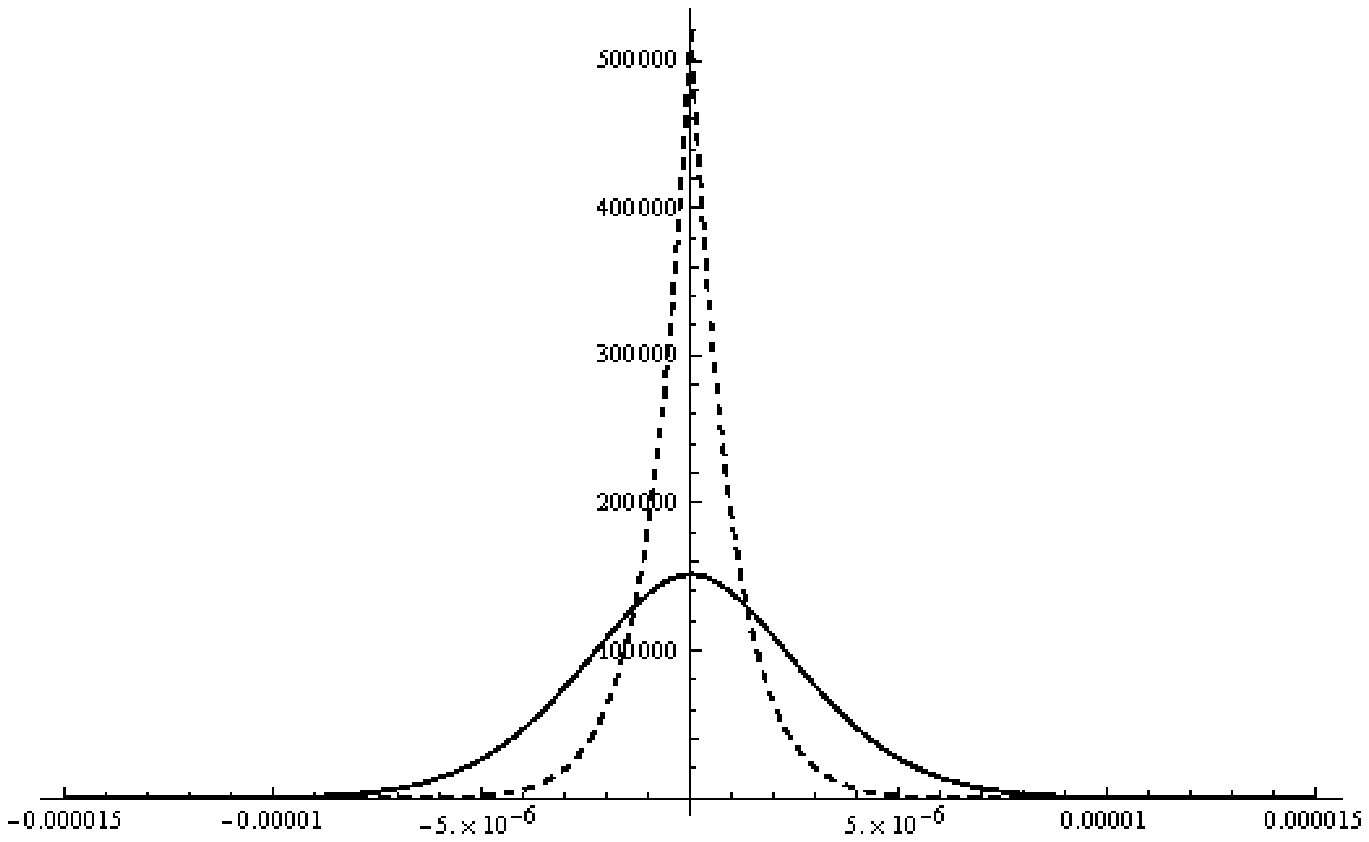}
\caption{Comparison of  density profiles of a BEC in a harmonic trap
with a BEC in a harmonic trap decorated with a delta function
($\Lambda=4.6$). The solid curve is the density profile of the BEC
in decorated potential. The dashed curve is the density profile of
the 1D harmonic trap ($\Lambda=0$). The parameter z is dimensionless
length defined after Equation (\ref{Scheq}). The other parameters
are the same as Fig. \ref{critemp}.} \label{Denprof}
\end{figure}

Finally, we compare density profiles of condensates for a linear
trap and a linear trap decorated with a delta function ($\Lambda
=4.6$) in Fig. (\ref{Denprof}). Since the ground state wave
functions can be calculated analytically for both cases, we find the
density profiles by taking the absolute square of the ground state
wave functions.


\section{Conclusion}

We have investigated the effect of the tight dimple potential on the
linearly confined one dimensional BEC. We model the dimple potential
with the Dirac \del function. In our model system, the increase in
the strength of the Dirac \del function can be interpreted as
increasing the depth of a dimple potential. This allows for
analytical expressions for the eigenfunctions of the system and a
simple eigenvalue equation greatly simplifying numerical treatment.
We have calculated the critical temperature, the chemical potential
and the condensate fraction and demonstrated the effect of the
dimple potential. We have found that the critical temperature can be
enhanced by an order of magnitude for experimentally accessible
dimple potential parameters. In general, we find that $T_c$
increases with the relative strength  for an attractive dimple and
decreases for repulsive dimple for a non-interacting gas. However as
our results show the decrease of the critical temperature for the
negative dimple is not as rapid as the increase for the repulsive
case. These results may be used to propose that the dimple
potentials can be a good candidate to circumvent the disadvantage of
the simple quadrupole trap mentioned in Reference \cite{dilute}.
Repulsive dimple can be used to repel atoms from the vicinity of the
node ($x=0$) of the quadrupole trap. Moreover, it may be also
possible to use attractive dimple potentials for the same reason
because only dimple type potentials are also able to trap several
atoms \cite{ma}. So they can prevent loosing atoms in the vicinity
of the node. Moreover the increase of the critical temperature will
be a further advantage for attractive dimple potentials.

We have also analyzed the change of the condensate fraction with
respect to the strength of the Dirac \del function  at a constant
temperature ($T=T_c^0$), and with respect to temperature at a
constant strength. It has been shown that the condensate fraction
can be increased considerably and large condensates can be achieved
at higher temperatures due to the strong localization effect of the
dimple potential. Finally, we have determined and compared the
density profiles of the linear trap and the decorated trap with the
Dirac \del function at the equilibrium point using analytical
solutions of the model system. Comparing the graphics of density
profiles, we see that a dimple potential maintain a considerably
higher density at the center of the linear trap.

The comparison of the critical temperature values for BEC' s in
linear trap with a dimple and harmonic potential with a dimple
shows that the critical temperature values are higher for harmonic
trap for the same dimple strength which is an expected result
since the confinement of harmonic potential are more powerful.
However the simplicity of the linear trap may still be useful in
obtaining BEC.

We have also presented a semi-classical method for calculating
various quantities such as entropy, critical temperature and
condensate fraction. The results show that as the dimple strength
increases the semi-classical approximation gives better results for
non-interacting gas.

We believe that the presented results  obtained for the
noninteracting condensate in a quadrupole trap with a dimple by
modeling dimple potentials by using a $\delta$ function can provide
a theoretical model for such experiments.


\ack

We acknowledge support by TUBITAK (Project No:108T003). The authors thank to
J. Armijo  for fruitful discussion.



\section*{References}
\begin{thebibliography}{44}

\bibitem{anderson} M. H. Anderson, J. R. Ensher, M. R. Matthews, C. E. Wieman and  E.
A. Cornell Science {\bf 269} (1995) 198.

\bibitem{davis}  K. B. Davis,  M. O. Mewes,  M. R. Andrews,  N.
J. Van Druten, D. S. Durfee, D. M. Kurn and  W. Ketterle  \PRL {\bf
75} (1995) 3969.

\bibitem{bradley}  C. C. Bradley, C. A. Sackett, J. J. Tollett and
R. G. Hulet \PRL {\bf 75} (1995) 1687.

\bibitem{khawaja} U. Al Khawaja, J. O. Andersen, N. P. Proukakis
and H. T. C. Stoof {\it Phys. Rev. A} {\bf 66} (2002) 013615.

\bibitem{ketterle} W. Ketterle and  N. J. Van Druten  {\it Phys. Rev. A}  {\bf 54} (1996) 656.

\bibitem{druten} N. J. Van Druten and W. Ketterle \PRL {\bf 79} (1997)
549.

\bibitem{gorlitz}  A. G\"orlitz, J. M. Vogels, A. E. Leanhardt, C. Raman,
T. L. Gustavson, J. R. Abo-Shaeer, A. P. v, Gupta S, S. Inouye, T.
Rosenband and W. Ketterle \PRL {\bf 87} (2001) 130402.

\bibitem{ott}  H. Ott, J. Fortagh, G. Schlotterbeck, A. Grossmann and
C. Zimmermann \PRL {\bf 87} (2001) 230401.

\bibitem{hansel}  W. H\"ansel, P. Hommelhoff, T. W. Hansch and J. Reichel Nature {\bf
413} (2001) 501.

\bibitem{schreck}  F. Schreck, L. Khaykovich, K. L. Corwin, G. Ferrari, T. Bourdel,
 J. Cubizolles and C. Salomon \PRL {\bf 87} (2001) 080403.

\bibitem{bruce} G. D. Bruce, S. L. Bromley, G. Smirne, L. Torralbo-Campo, and D. Cassetari  {\it Phys. Rev. A}  {\bf 84} (2011) 053410.

\bibitem{jacqmin} T. Jacqmin, B. Fang, T. Berrada, T. Roscilde, and I. Bouchoule  {\it Phys. Rev. A}  {\bf 86} (2012) 043626.

\bibitem{yukalov} V. I. Yukalov  {\it Phys. Rev. A}  {\bf 72} (2005) 033608.

\bibitem{armijo} J. Armijo, T. Jacqmin, K. Kheruntsyan, and I. Bouchoule  {\it Phys. Rev. A}  {\bf 83} (2011) 021605.

\bibitem{bouchoule} I. Bouchoule, K. Kheruntsyan, and G. V. Shlyapnikov  {\it Phys. Rev. A}  {\bf 75} (2007) 031606(R).

\bibitem{cavalcanti} R. M. Cavalcanti, P. Giacconi, G. Pupillo and
R. Soldati Phys. Rev. A {\bf 65} (2002) 053606.

\bibitem{pinkse} P. W. H. Pinkse, A. Mosk, M. Weidem\"{u}ller, M. W. Reynolds, T. W. Hijmans and
J. T. M. Walraven \PRL  {\bf 78} (1997) 990.

\bibitem{utem}  H. Uncu,  D. Tarhan, E. Demiralp and O. E. Mustecaplioglu Phys. Rev. A {\bf
76} (2007) 013618.

\bibitem{kurn} D. M. Stamper-Kurn, H.J. Miesner, A. P. Chikkatur, S. Inouye,
J. Stenger and  W. Ketterle \PRL  {\bf 81} (1998) 2194.

\bibitem{garrett}  M. C. Garrett \textit{et all.} Phys. Rev. A {\bf
83} (2011) 013630.

\bibitem{ranjani} S. Ranjani, U. Roy, P. K. Panigrahi and A. K. Kapoor J. Phys. B: At. Mol. Opt. Phys. {\bf
41} (2008) 235301.

\bibitem{busch}  J. Goold, D. Donoghue and T. Busch J. Phys. B: At. Mol. Opt. Phys. {\bf
41} (2008) 215301.

\bibitem{dilute}  C. J. Petchick and H. Smith {\it Bose-Einstein
Condensation in Dilute Gases} (Cambridge University Press,
Cambridge) (2001).

\bibitem{lighthill} M. J. Lighthill An introduction to
Fourier analysis and generalized functions (Cambridge: University
Press) (1959).

\bibitem{wangxin} WANG Xin, TANG Liang-Hui, WU Reng-Lai, WANG Nan, WANG Nan, LIU Quan-Hui Commun. Theor. Phys {\bf
53} (2010) 247.

\bibitem{landau} L. D. Landau and  E. M. Lifshitz Quantum Mechanics (Oxford: Buuterworth
Heinemann) (1977).

\bibitem{schwinger} J. Schwinger Quantum Mechanics (Berlin: Springer
Verlag) (2001).

\bibitem{stringari}  L. Pitaevskii, S. Stringari {\it Bose-Einstein
Condensation} (Oxford: Clarendon Press) (2003).

\bibitem{ma} Z. H. Ma, J. F. Christopher and Z. L. Cornish \JPB {\bf
37} (2004) 3187.


\end{thebibliography}
\end{document}